\documentstyle[12pt,epsfig]{article}
\def\p2t{p_{2\bot}}
\def\q2t{q_{2\bot}}
\begin{document}
\rightline{LUNFD6/(NFFL-7207) 2001}
\rightline{December 2001}          
\def\emline#1#2#3#4#5#6{%
       \put(#1,#2){\special{em:moveto}}%
       \put(#4,#5){\special{em:lineto}}}
\begin{center}
\vspace{1cm}
{\Large\bf
 Heavy Quark Production at the TEVATRON\\[5mm]
 in the  Semihard QCD Approach \\[5mm]
and the Unintegrated Gluon Distribution} \\[5mm]
\vspace{1cm}
{\bf    A.V. Lipatov}
\footnote{\parbox[t]{10cm}{Electronic address: artem$_-$lipatov@mail.ru}}\\
{\sl Department of Physics,~Moscow~State~University,\\
119899~Moscow,~Russia}\\
\vspace{0.2cm}
{\bf V.A. Saleev}
\footnote{\parbox[t]{10cm}{Electronic address: saleev@info.ssu.samara.ru}}\\
{\sl Department of Physics, Samara~State~University,\\
443011~Samara,~Russia}\\
\vspace{0.2cm}
{\bf N.P.~Zotov}
  \footnote{\parbox[t]{10cm}{Electronic address: 
zotov@theory.sinp.msu.ru}}\\
{\sl Skobeltsyn~Institute~of~Nuclear~Physics,\\
~Moscow~State~University,
~119899~Moscow,~Russia}\\
  \end{center}
 \vspace*{1cm}~\\

\begin{abstract}
Processes of heavy quark production at TEVATRON energies
are considered using the semihard
($k_T$ factorization) QCD approach with emphasis
of the BFKL dynamics of gluon distributions. We investigate the dependence
 of the $p_T$ distribution of   heavy quark 
production (presented in the form of integrated cross-sections)
 on different forms of the  unintegrated gluon distribution.
The theoretical results are compared with recent 
D0 and CDF  experimental data on beauty 
production.
\end{abstract}

\newpage
\noindent
\section {Introduction}

It is very well known that from heavy ($c-$ and $b-$) quark  
production
processes one can obtain unique information on gluon structure function
of the proton  and the three gluon QCD vertex because of the dominance of 
the gluon-gluon fusion subprocess in the framework of QCD.
These issues are important for physics at future colliders
(such as LHC): many processes at these colliders will be determined
by small $x$ gluon distributions.  

Recently D0 and CDF Collaborations  have reported~\cite{r1,r2} new
experimental data on the cross section of inelastic
open beauty production at the TEVATRON energies. A comparision of these 
results with 
NLO pQCD calculations shows that the calculations  underestimate the cross section
by a factor of $\sim 2$. Therefore, it would be certainly
 reasonable to try a different way.


In the energy range of modern colliders (HERA, TEVATRON and LHC) the 
interaction dynamics is governed by the 
properties of parton distributions in the small $x$ region. This domain is 
characterized by the double inequality $s\gg\mu^2\simeq\hat s\gg\Lambda^2$,
which shows that the typical parton interaction scale $\mu$ (mass $m_c$
or $p_T$ of heavy quark) is much higher
than the QCD parameter $\Lambda$, but is much lower than the total 
center of mass 
energy $\sqrt s$. The situation is therefore classified as ``semihard'':
the processes occur in small $x$
region ($x\simeq M^2/s\ll 1$) and the cross sections of heavy
quark production processes are determined by the   
behavior of gluon distributions in this region.

It is known also that in the small $x$ region
 the  factorization of QCD
 subprocess cross section and parton structure functions in a hadron
assumed by the standard parton model
is broken \cite{r3}-\cite{r7}.
The resummation \cite{r3,r4,r6,r7} of the terms
$[\mbox{ln}(\mu^2/\Lambda^2)\,\alpha_s]^n$,
$[\mbox{ln}(\mu^2/\Lambda^2)\,\mbox{ln}(1/x)\,\alpha_s]^n$ and
$[\mbox{ln}(1/x)\,\alpha_s]^n$ in the semihard ($k_T$-factorization)
approach (SHA) of QCD results 
in the unintegrated gluon
distributions $\Phi(x,q_{T}^2,\mu)$, which determine the probability 
to find a gluon  carrying the longitudinal momentum fraction 
$x$
and transverse momentum $q_T$ at the probing scale $\mu^2$.
They obey the BFKL 
equation~\cite{r4} and reduce to the conventional gluon density
once the $q_{T}$ dependence is integrated out ($\mu^2 \equiv Q^2$):
\begin{equation} \label{kt}
\int_0^{Q^2}\!\!\Phi(x, q_{T}^2, Q^2_0)\;dq_{T}^2=
x\,G(x, Q^2).
\end{equation}

To calculate the cross section of a physical process, the unintegrated 
functions $\Phi_i$  have to be convoluted
with off-mass shell matrix elements
corresponding to the relevant partonic subprocesses.
In the off-mass shell matrix element the virtual gluon polarization 
tensor is taken in the form of the SHA prescription \cite{r3,r4}:
\begin{equation}
L^{(g)}_{\mu\nu}=\overline{\epsilon_2^{\mu}\epsilon_2^{*\nu}}
  =p^\mu p^\nu x^2/|q_T|^2 
  =q_T^\mu q_T^\nu/|q_T|^2
\end{equation}

Nowadays, the significance of the $k_T$ factorization (semihard) approach 
becomes more and more
 commonly recognized. Its applications to a variety of photo-, lepto- and
 hadroproduction processes are widely discussed in the literature
 \cite{r5},\cite{r8}-\cite{r11}. Remarkable agreement is
 found between the data and the theoretical calculations regarding the
 photo-~\cite{r12} and electroproduction \cite{r13,r14} of $D^*$ mesons,
 forward jets \cite{r15, r16},
 as well as for specific kinematic correlations observed in the
 associated $D^*${+}jets photoproduction \cite{r17} at HERA and also
 the hadroproduction of beauty \cite{r18,r19}, $\chi_c$ 
and $J/\Psi$ \cite{r20,r21} at the Tevatron.
 The theoretical predictions made in ref. \cite{r22} has triggered
 a dedicated experimental analysis \cite{r23} of the $J/\Psi$ polarization
 (i.e., spin alignement) at HERA conditions.
   
  Here study of heavy quark production processes
in $p\bar p-$collisions at TEVATRON with the SHA is presented.
 Similar investigations have been done by many
authors~\cite{r5,r18,r19} earlier. But in the recent papers
~\cite{r18,r19} obtained results contradict each other.
 
 We investigate the sensitivity of the inelastic 
cross section
of heavy quark production processes at TEVATRON to
different uintegrated gluon distributions. Special attention is given to
the unintegrated gluon distributions obtained from BFKL evolution
equation.  The outline of our paper is the following. In sect. 2, we give
the formulas for the cross sections of heavy quark hadroproduction
in the SHA of QCD. Then, in sect. 3, we describe the unintegrated 
distributions which we use for our calculations. In sect. 4 we give 
the matrix elements for gluon-gluon fusion QCD subprocesses. 
In sect. 5, we present 
the results of our calculations. Finally, in sect. 6, we give some 
conclusions\footnote{The obtained results have been reported at Workshop
DIS 2001, Bologna, 27 April - 1 May 2001.}.

\section{SHA QCD Cross-Section for Heavy Quark Production in $p\bar p$ Collisions}

We calculate the total and differential cross sections of heavy quark
 production $p\bar p\to Q\bar Q X$
via the gluon-gluon fusion QCD subprocess (Fig.1) in the framework
of the SHA.
\begin{figure}
\begin{center}
\epsfig{figure= 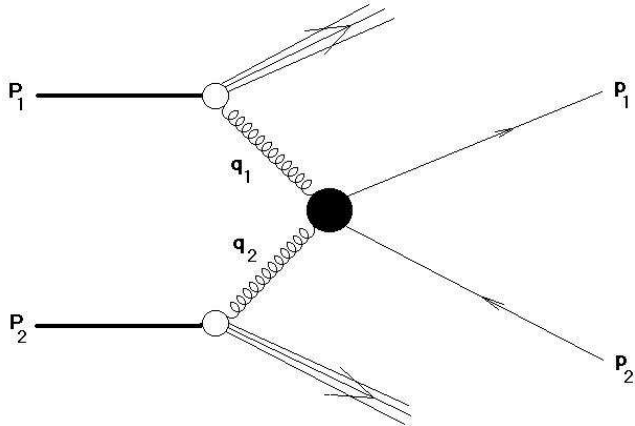,width=10cm,height=7cm}
\end{center}
\caption{
 QCD diagram for the open heavy quark production.}
\label{fig1}
\end{figure}
First of all we take into account the transverse momentum of initial 
gluons
($\vec{q}_{1,2T}$), their virtualities ($q^2_{1,2}=-\vec{q}^2_{1,2T}$) 
and the alignment of gluon polarization vectors along their transverse momenta 
given by (2)~\cite{r3,r4,r6,r7}.
Let us define Sudakov variables of the process $p\bar p\to Q\bar Q X$
(Fig.1):
\begin{eqnarray}
p_1 = \alpha_1 P_1+\beta_1 P_2+p_{1T},\,\,\,
p_2 = \alpha_2 P_1+\beta_2 P_2+p_{2T}\nonumber\\
q_1 = x_1P_1+q_{1T},\,\,\,
q_2 = x_2P_2+q_{2T},
\end{eqnarray}
where
$$p_1^2=p_2^2=M^2,\,\,\,q_1^2=q_{1T}^2,\,\,\,q_2^2=q_{2T}^2,$$
$p_1$, $p_2$ and $q_1$, $q_2$  are 4-momenta of the heavy quarks 
and the gluons respectively, $p_{1T},~~p_{2T}$ and
 $q_{1T},~~q_{2T}$ are transverse 4-momenta of quarks and gluons.
 In the center of mass frame of colliding particles we can write
 $P_1=(E,0,0,E)$, $P_2=(E,0,0,-E)$, where $E=\sqrt s/2$, $P_1^2=P_2^2=0$
  and $(P_1P_2)=s/2$.
Sudakov variables are expressed  as follows:
\begin{eqnarray}
\alpha_1 &=&\frac{M_{1T}}{\sqrt s}\exp(y_1^\ast),\,\,\,
\alpha_2 =\frac{M_{2T}}{\sqrt s}\exp(y_2^\ast),\nonumber\\
\beta_1 &=&\frac{M_{1T}}{\sqrt s}\exp(-y_1^\ast),\,\,\,
\beta_2 =\frac{M_{2T}}{\sqrt s}\exp(-y_2^\ast),
\end{eqnarray}
where $M_{1,2T}^2=M^2+p_{1,2T}^2$, $y_{1,2}^{\ast}$ are rapidities
of heavy quarks, M is heavy quark mass.
From conservation laws we can easly obtain the following relations:
\begin{eqnarray}
q_{1T}+q_{2T}=p_{1T}+p_{2T},\,\,\,
 x_1=\alpha_1 +\alpha_2,\,\,\,
 x_2=\beta_1 +\beta_2
\end{eqnarray}
The differential cross section of heavy quark  production has the
following form:
\begin{eqnarray}
 d\sigma(p\bar p\to Q\bar Q\,X) =\frac{dx_1} 
{x_1}\,\Phi(x_1,\,q_{1T}^2,\,Q^2)\,\frac{d\phi_1}{2\pi}\,dq_{1T}^2\,\times 
\atop
\times\,\frac{dx_2} 
{x_2}\,\Phi(x_2,\,q_{2T}^2,\,Q^2)\,\frac{d\phi_2}{2\pi}\,d{q_{2T}^2}\,d\hat 
\sigma(g^*g^*\to Q\bar Q),
\end{eqnarray}
where $\Phi(x,\,q_{T}^2,Q^2)$ is 
an unintegrated
gluon distribution in the proton, $d\hat \sigma(g^*g^*\to Q\bar 
Q)$ is the differential cross section of gluon-gluon fusion subprocess.
We used the following form:
\begin{equation}
d\hat \sigma(g^*g^*\to Q\bar Q) = {(2\pi)^4\over 2\hat s}\,\sum 
{|M|^2_{SHA}(g^*g^*\to Q\bar Q)}\,{d^3p_1\over (2\pi)^3\,2p_1^0}\,
{d^3p_2\over (2\pi)^3\,2p_2^0}\,\delta^{(4)}(q_1 + q_2 - p_1 - p_2),
\end{equation}
where $\sum {|M|^2_{SHA}(g^*g^*\to Q\bar Q)}$ is the  off mass shell 
matrix element and
\begin{equation}
\delta^{(4)}(q_1 + q_2 - p_1 - p_2) = \delta(q_1^0 + q_2^0 - p_1^0 - 
p_2^0)\,\delta(\vec q_{1T} + \vec q_{2T} - 
\vec p_{1T} - \vec p_{2T})\,\delta(q_1^3 + 
q_1^3 - p_1^3 - p_2^3).
\end{equation} 
 In (7) $\sum$ indicates an averaging over polarizations
of initial gluons and a sum over polarizations of final quarks.
We have also
\begin{equation}
{d^3p_1\over (2\pi)^3\,2p_1^0}\,{d^3p_2\over (2\pi)^3\,2p_2^0} = 
{d^2p_{1T}\over 2(2\pi)^3}\,dy_1^*\,{d^2p_{2T}\over 2(2\pi)^3}\,dy_2^*.
\end{equation} 
 Integrating out the $p^2_{2T}, x_1$ and $x_2$ dependences in (6) and 
(9) with accounting of (7) and (8) we obtain the following formula
for the differential cross section of the process $p\bar p \to
Q\bar Q X$ in the framework of the SHA:
\begin{equation}
 d\sigma(p\bar p\to Q\bar Q\,X) =\frac{1}{16\pi 
(x_1\,x_2\,s)^2}\,\Phi(x_1,\
,q_{1T}^2,\,Q^2)\,\Phi(x_2,\,q_{2T}^2,\,Q^2)\,\times \atop
 \times \sum {|M|^2_{SHA}(g^*g^*\to Q\bar Q)}\,dy_1^*\,dy_2^*\,
dp_{1T}^2\,dq_{1T}^2\,dq_{2T}^2\,\frac{d\phi_1}{2\pi}
\frac{d\phi_2}{2\pi}\,\frac{d\phi_3}{2\pi}.
\end{equation}
 
 If we take the limit $q_{1T}\to 0$, $q_{2T}\to 0$ and if we average (10)
 over the transverse directions of the vectors $\vec q_{1T}$ and
 \ $\vec q_{2T}$, we obtain the formula for the the differential cross
 section of the process $p\bar p \to Q\bar Q X$ in the standard parton model:
\begin{equation}
d\sigma(p\bar p\to q\bar q\,X) = {x_1G(x_1,\,Q^2)\,x_2G(x_2,\,Q^2)\over
 16\pi\,(x_1 x_2 s)^2}\,
\sum {|M|_{PM}^2(gg\to Q\bar Q)}\,dy_1^*\,dy_2^*\,dp_{1T}^2,
\end{equation}
where $\sum {|M|^2_{PM}(gg\to Q\bar Q)}$ is matrix elements of 
gluon-gluon fusion QCD subprocess in the standard parton model 
(SPM).
Here $\sum$ idicates averaging over polarizations of on-shell
initial gluons and a sum over polarizations of final quarks.
We average over the transverse directions of $\vec q_{1T}^2$ 
and $\vec q_{2T}^2$ using the following expression: 
\begin{equation}
 \int dq_{1T}^2\,\int {\frac{d\phi_1}{2\pi}}\,\Phi(x_1,\,q_{1T}
^2)\,\int dq_{2T}^2\,\int {\frac{d\phi_2} 
{2\pi}}\,\Phi(x_2,\,q_{2T}^2)\,\sum
 {|M|_{SHA}^2(g^*g^*\to Q\bar Q} = \atop 
 = x_1G(x_1,\,Q^2)\,x_2G(x_2,\,Q^2)\,\sum {|M|_{PM}^2(gg\to Q\bar Q)},
\end{equation}
where
 
\begin{equation}
\int\limits_0^{2\pi} {d\phi_{1,2}\over 2\pi}\,{q_{1,2T}^{\mu}\,q_{1,2T}^{\nu}\over 
\vec q_{1,2T}^2} = {1\over 2}\,g^{\mu\nu}
\end{equation}

\section{Unintegrated Gluon Distribution Functions}

Various parametrizations of the 
unintegrated gluon distribution used in calculations are
discussed below. First, as in the publication
~\cite{r8,r9}, we used the following phenomenological parametrization
(LRSS-parametrization)~\cite{r3,r5}:
\begin{equation}
\Phi (x,\vec{q}_T\,^2)=\Phi_0\frac{0.05}{0.05 + x}(1-x)^3f_1(x,\vec{q}_T\,^2),
\end{equation}
where    
\begin{eqnarray}
f_1(x,\vec{q}_T\,^2)= \cases{
1,                          & $\mbox{if}\, \vec{q}_T\,^2 \le q_0^2(x)$, \cr
(q_0^2(x)/\vec{q}_T\,^2)^2, & $\mbox{if}\, \vec{q}_T\,^2 > q_0^2(x)$}
\end{eqnarray}
with $q_0^2(x)=q_0^2+\Lambda^2\exp(3.56\sqrt{ln(x_0/x)})$,   
$q_0^2 = 2$ GeV$^2$, $\Lambda = 56$ MeV, $x_0=1/3$. The  value  of the
parameter $q_0^2(x)$ can be considered as a typical transverse
momentum  of the
partons in the parton cascade which leads to natural
infrared cut-off in semihard approach. The normalization  factor
 of the  structure  function  $\Phi(x,\vec q_T\,^2) (\Phi_0 
= 0.97$ mb)  was
obtained  in ~\cite{r5}, from the beauty production  in  $p\bar p$
at $Sp\bar pS$ energy\footnote{The experimental data~\cite{r24}
relate to relatively low energy for applicability of the SHA.}.
 The effective gluon distribution $xG(x,Q^2)$, which was obtained 
from eq. (1) with (14) and (15), increases at not very low
 $x (0.01< x< 0.15)$ as $ xG(x,Q^2)\sim x^{-\Delta}$,
  where $\Delta \approx 0.5$ corresponds to the QCD  pomeron singularity
  given by summation of leading logarithmic contributions
($\alpha_s ln\frac{1}{x})^n$~\cite{r4}. This increase continues up to
$x=x_0$, where $x_0$ is a solution of the equation $q_0^2(x_0) = Q^2$.
 In the region $x<x_0$, there is saturation of the gluon
distribution function: $xG(x,Q^2)\approx \Phi_0 Q^2$ (curves 3 in Fig.2). 

Another parametrization is based on the numerical solution of the BFKL
evolution equations (RS--parametrization), which
 has the following form~\cite{r25}:
\begin{eqnarray}
\Phi(x, q^2) ={1\over 4\sqrt 2\,\pi^3}\,\frac{a_1}{a_2 + a_3 +a_4}[a_2 + a_3 
(\frac{Q_0^2}{q^2})
+ (\frac{Q_0^2}{q^2})^2 +
\nonumber \\
 \alpha x + 
 \frac{\beta}{\epsilon + \ln (1/x)}]
C_q [\frac{a_5}{a_5 + x}]^{1/2}[1 -
\nonumber \\
 a_6 x^{a_7}\ln{(q^2/a_8)}]
(1 + a_{11} x) (1  - x)^{a_9 + a_{10}\ln (q^2/a_8)},
\end{eqnarray}
where
\begin{equation}
C_q = \cases{
1,          & $\mbox{if}\, q^2  < q_0(x)$, \cr
q_0(x)/q^2, & $\mbox{if}\, q^2  > q_0(x)$.}
\end{equation}
 All parameters (see~\cite{r25}) $(a_1 - a_{11}, \alpha, \beta$ and
$\epsilon)$ were found by minimization of the differences between
left hand and right-hand of the BFKL-type equation for 
unintegrated gluon distribution $\Phi (x, q^2)$ at $Q_0^2 =$4 GeV$^2$.

\begin{figure}
\begin{center}
\epsfig{figure= 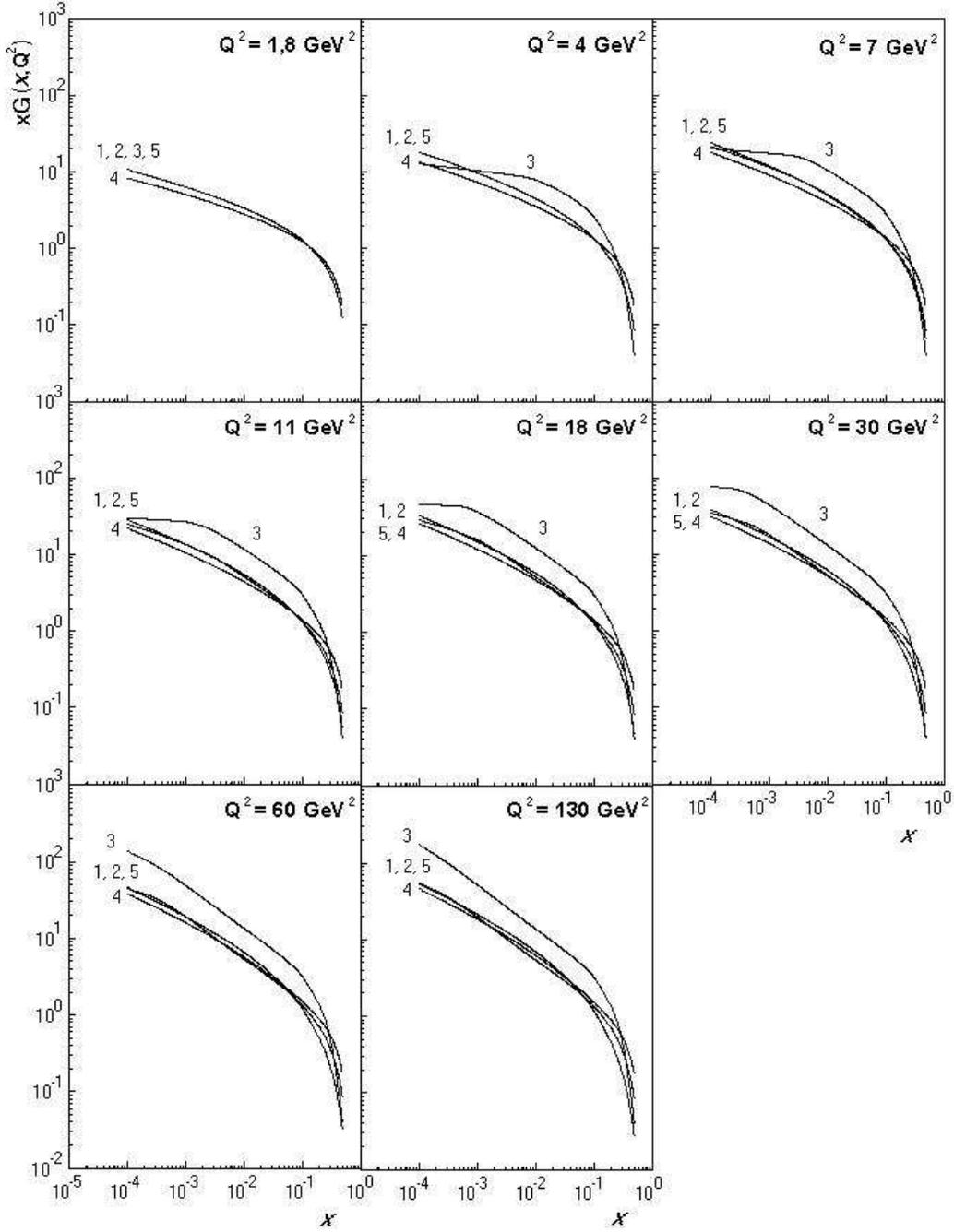,width=15cm,height=20cm}
\end{center}
\caption{
Effective gluon densities $xG(x,Q^2)$ obtained from different
unintegrated gluon distribution. Curve 1 correspond to the GRV 
collinear gluon density~\cite{r27}, curves 2, 3, 4, 5 correspond to
RS~\cite{r25} (at $Q_0^2 = 4$ GeV$^2$), LRSS~\cite{r5} (at $Q_0^2 = 2$ 
GeV$^2$), BFKL~\cite{r26} (at $Q_0^2 = 1$ GeV$^2$) parametrizations
and differential (at $Q_0^2 = 1$ GeV$^2$) gluon  
distribution.}  
\label{fig2}
\end{figure}

Then we  use the results of a BFKL-like
 parameterization\footnote{ Of course LRSS  and RS parametrizations
are BFKL - type too.} of
the unintegrated gluon distribution 
$\Phi(x,q^2_{T},\mu^2)$, according to the prescription given in~\cite{r26}. 
The proposed method lies upon a straightforward perturbative solution of
the BFKL equation where the collinear gluon density $x\,G(x,\mu^2)$
from the standard GRV set~\cite{r27} is used as
the boundary condition in the integral (1).
Technically, the unintegrated gluon density is calculated as a convolution
of collinear gluon density $G(x,\mu ^2)$
with universal weight factors~\cite{r26}:
\begin{equation} 
 \Phi(x,q_T^2,\mu^2) = \int_x^1
 {\cal G}(\eta,q_T^2,\mu^2)\,
 \frac{x}{\eta}\,G(\frac{x}{\eta},\mu^2)\,d\eta,
\end{equation}
where
\begin{equation} 
 {\cal G}(\eta,q_T^2,\mu^2)=\frac{\bar{\alpha}_s}{\eta q_T^2}\,
 J_0(2\sqrt{\bar{\alpha}_s\ln(1/\eta)\ln(\mu^2/q_T^2)}),
 \qquad q_T^2<\mu^2,
\end{equation}

\begin{equation}
 {\cal G}(\eta,q_T^2,\mu^2)=\frac{\bar{\alpha}_s}{\eta q_T^2}\,
 I_0(2\sqrt{\bar{\alpha}_s\ln(1/\eta)\ln(q_T^2/\mu^2)}),
 \qquad q_T^2>\mu^2,
\end{equation}
where $J_0$ and $I_0$ stand for Bessel functions (of real and imaginary
arguments, respectively), and $\bar{\alpha}_s=3{\alpha}_s/\pi$.
The  parameter $\bar{\alpha}_s$ is connected with
 the Pomeron trajectory intercept:
$\Delta=\bar{\alpha}_s4\ln{2}$ in the LO and 
$\Delta=\bar{\alpha}_s4\ln{2}-N\bar{\alpha}_s^2$ in the NLO approximations,
respectively, where $N \sim 18$  \cite{r28}. The latter value 
of $\Delta$ have dramatic 
consequences for high energy phenomenology~\cite{r29}.
However some resummation procedures proposed in the last years
lead to positive value of $\Delta (\sim 0.2 - 0.3)$~\cite{r30,r31}.
Therefore in our calculations with (18) we used only
the solution of LO BFKL equation and 
considered $\Delta$ as a free parameter varying it from 0.166 to 0.53.
Pomeron intercept parameter $\Delta = 0.35$ was obtained
from the description of $p_T$ spectrum
of $D^*$ meson electroproduction at HERA~\cite{r12}.
 We used this value of the the parameter $\Delta$ in 
present paper.

 Finally we tried the so called differential unintegrated gluon structure
function of a proton, which could be obtained  by differentiation
of the collinear gluon density $xG(x,Q^2)$~\cite{r3,r4,r32},
 for example one from the standard GRV set~\cite{r27}, according to (1)

\begin{equation}
\Phi(x,q_T^2,Q^2) = {d{\,xG(x,Q^2)} \over d{\,\ln Q^2}}\bigg\vert _{Q^2 = 
q_T^2}
\end{equation}

 In  expression
(1) the accounting of the contribution of the
unintegrated gluon distribution to integral at low $q_T^2$ 
region ($0 < q_T^2 < Q_0^2$) was done by the collinear gluon
density $xG(x,Q_0^2)$, where $Q_0^2 = (1\div 4)$\,GeV$^2$ in dependence on
different parametrizations of unintegrated gluon distribution.

   Fig. 2 shows the $x$ dependence of the effective gluon density
$xG(x,Q^2)$ at fixed values of $Q^2$ obtained with help of
 the definition (1) for different parametrizations of unintegrated
gluon distribution discussed above. The curve (1) corresponds
 to the standard 
GRV parametrization of collinear gluon distribution function.
Curve 2, 3, 4, 5 correspond to effective gluon densities obtained
from unintegrated gluon distribution functions $\Phi(x,q_T^2)$
with help of the definition (1) for RS, LRSS, BFKL  parametrizations 
and the  definition (21) at different values of  $Q_0^2$ parameter.

 The effective gluon density $xG(x,Q^2)$ obtained for the LRSS
parametrization  of unintegrated gluon distribution  $\Phi(x,q_T^2)$ 
 differs strongly from other gluon densities normalizated to 
GRV collinear gluon density. It is result of the normalization of the 
LRSS parametrization of $\Phi(x,q_T^2)$ obtained from the $Sp\bar pS$
experinental data for $b\bar b$ production in $p\bar p$ collisions
~\cite{r24}.

\section{ Matrix Elements for Gluon-Gluon Fusion QCD \\
Subprocesses}

\begin{figure}
\begin{center}
\epsfig{figure= 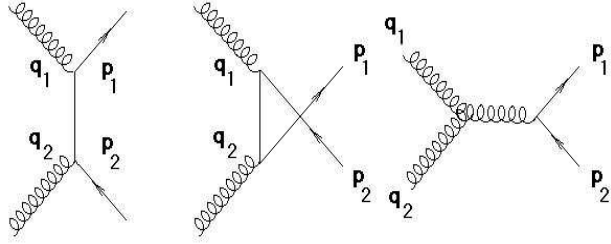,width=10cm,height=7cm}
\end{center}
\caption{ QCD diagramms of  gluon-gluon fusion subprocesses
for heavy quark production.}
\label{fig3}
\end{figure}
The subprocess  $g^*g^*\to Q\bar Q$ is described in the LO pQCD
by three diagramms shown in Fig. 3. In according with the standard
Feynman rules for QCD diagramms: 
 
\begin{eqnarray}
 M_A = \bar u(p_1)\,(-ig\gamma^{\mu})\,\varepsilon_{\mu}(q_1)\,   
i{\hat p_1 - \hat q_1 + m\over (p_1 - q_1)^2 - m^2}\,
(-ig\gamma^{\nu})\,\varepsilon_{\nu}(q_2)\,v(p_2), \nonumber \\ 
 M_B = \bar u(p_1)\,(-ig\gamma^{\nu})\,\varepsilon_{\nu}(q_2)\,
i{\hat p_1 - \hat q_2 + m\over  (p_1 - q_2)^2 - m^2}\,
(-ig\gamma^{\mu})\,\varepsilon_{\mu}(q_1)\,v(p_2), \nonumber \\ 
 M_C = \bar u(p_1)\,C^{\mu\nu\lambda}
(-q_1,-q_2,q_1+q_2)\,{g^2\,\varepsilon_{\mu}(q_1)\,
\varepsilon_{\nu}(q_2) \over (q_1 + q_2)^2}\,\gamma_{\lambda}\,v(p_2),
\end{eqnarray}

where $\varepsilon_{\mu}(q_1)$ and $\varepsilon_{\nu}(q_2)$ are the
polarization vectors of the initial gluons, 
$C^{\mu\nu\lambda}(q_1,q_2,q_3)$ is the standard QCD
 three gluon  vertex:

\begin{equation}
C^{\mu\nu\lambda}(q_1,q_2,q_3)=i((q_2 - q_1)^\lambda)\,g^{\mu\nu} + 
 (q_3 - q_2)^{\mu}\,g^{\nu\lambda} + (q_1 - q_3)^{\nu}\,g^{\lambda\mu}).
\end{equation}

The Mandelstam variables for gluon-gluon fusion subprocess
$g^*g^* \to Q\bar Q$ are 

\begin{eqnarray}
\hat s = (q_1 + q_2)^2 = (p_1 + p_2)^2, \nonumber \\ 
\hat t = (q_2 - p_2)^2 = (p_1 - q_1)^2, \nonumber \\ 
\hat u = (q_1 - p_2)^2 = (p_1 - q_2)^2,
\end{eqnarray}
 
and 

\begin{equation}
 \hat s + \hat t + \hat u = 2\,M^2 + q_{1T}^2 + q_{2T}^2.
\end{equation}

Let present the matrix elements in the following forms:

\begin{equation}
M_{A,B,C} = 
\varepsilon_{\mu}(q_1)\,\varepsilon_{\nu}(q_2)\,M_{A,B,C}^{\mu\nu},
\end{equation}

then averaging over initial gluon polarizations 
and summing over final quark polarizations  we have

\begin{equation}
 \sum {|M|^2} = C_A\sum{|M|_A^2} + C_B\sum{|M|_B^2} +
 C_C\sum{|M|_C^2} + \atop
 +\, 2\,C_{AB}\sum{|M|_{AB}^2} + 2\,C_{AC}\sum{|M|_{AC}^2} + 
 2\,C_{BC}\sum{|M|_{BC}^2},
\end{equation}
where

\begin{equation}
\sum{|M|_i^2} = {1\over 4}\,L_{\mu\alpha}^{(g)}(q_1)\,
L_{\nu\beta}^{(g)}(q_2)\,M_i^{\mu\nu}\,M_i^{*\,\alpha\beta}
\end{equation}

Here the index $i=A,\,B,\,C,\,AB,\,AC,\,BC$ and the color factors are 
$C_A = C_B = 1/12$,  $C_A = C_B = 1/12$,
$C_C = 3/16$, $C_{AB} = - 1/96$ ¨ $C_{AC} = C_{BC} = 3/32$ accordingly.
The effictive gluon polarization tensor $L_{\mu\nu}^{(g)}(q)$ is taken 
from (2).

The calculation of  $\sum{|M|_{SHA}^2}(g^*g^*\to Q\bar Q)$ was done 
analitically by REDUCE system. The obtained results coincide with 
the ones from ref.~\cite{r6}.

For the calculations of $\sum {|M|_{PM}^2(gg\to Q\bar Q)}$ in the
framework of the SPM we used the following form of
the gluon polarization tensor in axial gauge:

\begin{equation}
L^{\mu\nu}_{(g)}(q_1,\,q_2) = - g^{\mu\nu} + {q_1^{\mu}\,q_2^{\nu} + 
 q_1^{\nu}\,q_2^{\mu}\over (q_1,\,q_2)} - q_2^2\,
{q_1^{\mu}\,q_1^{\nu}\over (q_1,\,q_2)^2},
\end{equation}
Obtained expression for $\sum {|M|^2_{PM}(gg \to 
 Q\bar Q)}$ coincides with the results of ref.~\cite{r33}.
  
\section{Results of Calculations}

\begin{figure}
\begin{center}
\epsfig{figure= 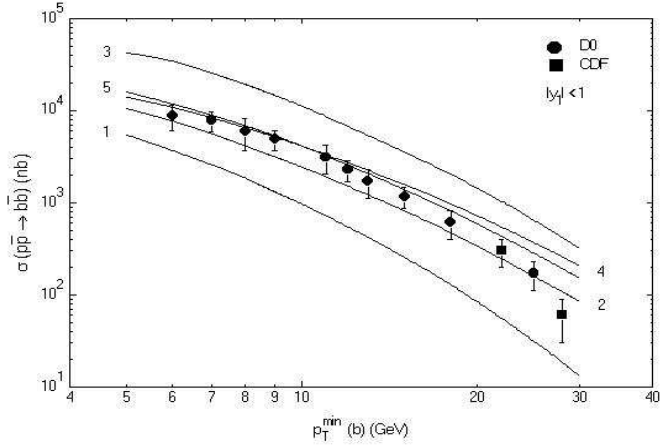,width=10cm,height=7cm}
\end{center}
\caption{
The $b\bar b$ production cross section at $\sqrt s$ = 1.8 TeV.
Curves 1 - 5 are the same as in Fig. 2.}
\label{fig4}
\end{figure}

The calculations of the heavy quark hadroproduction cross section
in the SHA have been made accoding to eqn. (10)  for  $\vec 
q_T\,^2 > q_0^2$ GeV$^2$. For $\vec q_T\,^2 \leq q_0^2$ GeV$^2$
 we set $|\vec q_T| = $ 0 in the 
matrix elements of subprocesses, take $\sum {|M|_{PM}^2(gg\to Q\bar Q)}$
instead of ${|M|_{SHA}^2(g^*g^*\to Q\bar Q)}$  
and use the eqn. (11) of the standard 
parton model (SPM).
\begin{figure}
\begin{center}
\epsfig{figure= 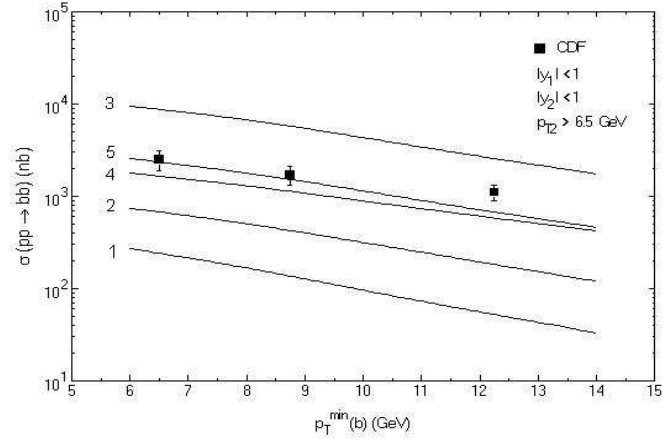,width=10cm,height=7cm}
\end{center}
\caption{
 The $b\bar b$ production cross section  at $\sqrt s$ = 1.8 TeV.
Curves 1 - 5 are the same as in Fig. 2.}
\label{fig5}
\end{figure}
 The choice of the critical value of parameter
$\vec q_T\,^2 = q_0^2 = (1\div 4)$ GeV$^2$ is determined by the requirement
of the small value of $\alpha _s(\vec q_T\,^2)$ in the region 
$\vec q_T\,^2 > (1\div 4)$ GeV$^2$, where in fact
 $\alpha _s(\vec q_T\,^2) \leq 0.26.$
\begin{figure}
\begin{center}
\epsfig{figure= 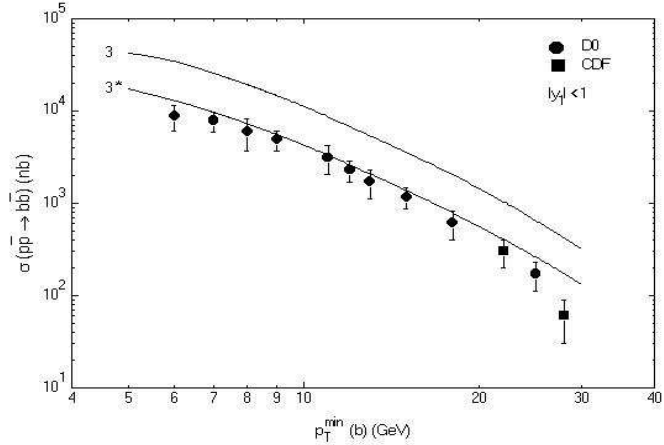,width=10cm,height=7cm}
\end{center}
\caption{
 The $b\bar b$ production cross secton at $\sqrt s$ = 1.8 TeV.
Curves 3 and 3$^*$ correspond to the SHA calculations with
the LRSS unintegrated gluon distribution at $m_b = 4.75$ GeV,
$\Lambda_{QCD}$ = 150 MeV, $Q_0^2 = 2$ GeV$^2$ and
 $m_b = 5.0$ GeV, $\Lambda_{QCD}$ = 100 MeV, $Q_0^2 = 1$ GeV$^2$
accordingly. }
\label{fig6}
\end{figure}

The results of our calculations for the cross sections of  
$b \bar b$  production at TEAVTRON 
 are shown in Figs. $4\div 8$. These figures show the  
 $p_{1T}^{min}$ dependence of the $b\bar b$
 production cross section at TEVATRON energies calculated
 with $m_b = 4.75$ GeV, $\Lambda_{QCD} = 150$ MeV and $|y_1^*| < 1$
 ( Fig. 4)~\cite{r1}, and 
$|y_1^*| < 1$, $|y_2^*| < 1$, $p_{2T} > 6.5$ GeV ( Fig. 5)~\cite{r2}
\footnote{The cross section does not depend on $p_{1T}^{min}$ in the
region 1 GeV$ \le p_{1T}^{min} \le 5$ GeV approximately.}.
\begin{figure}
\begin{center}
\epsfig{figure= 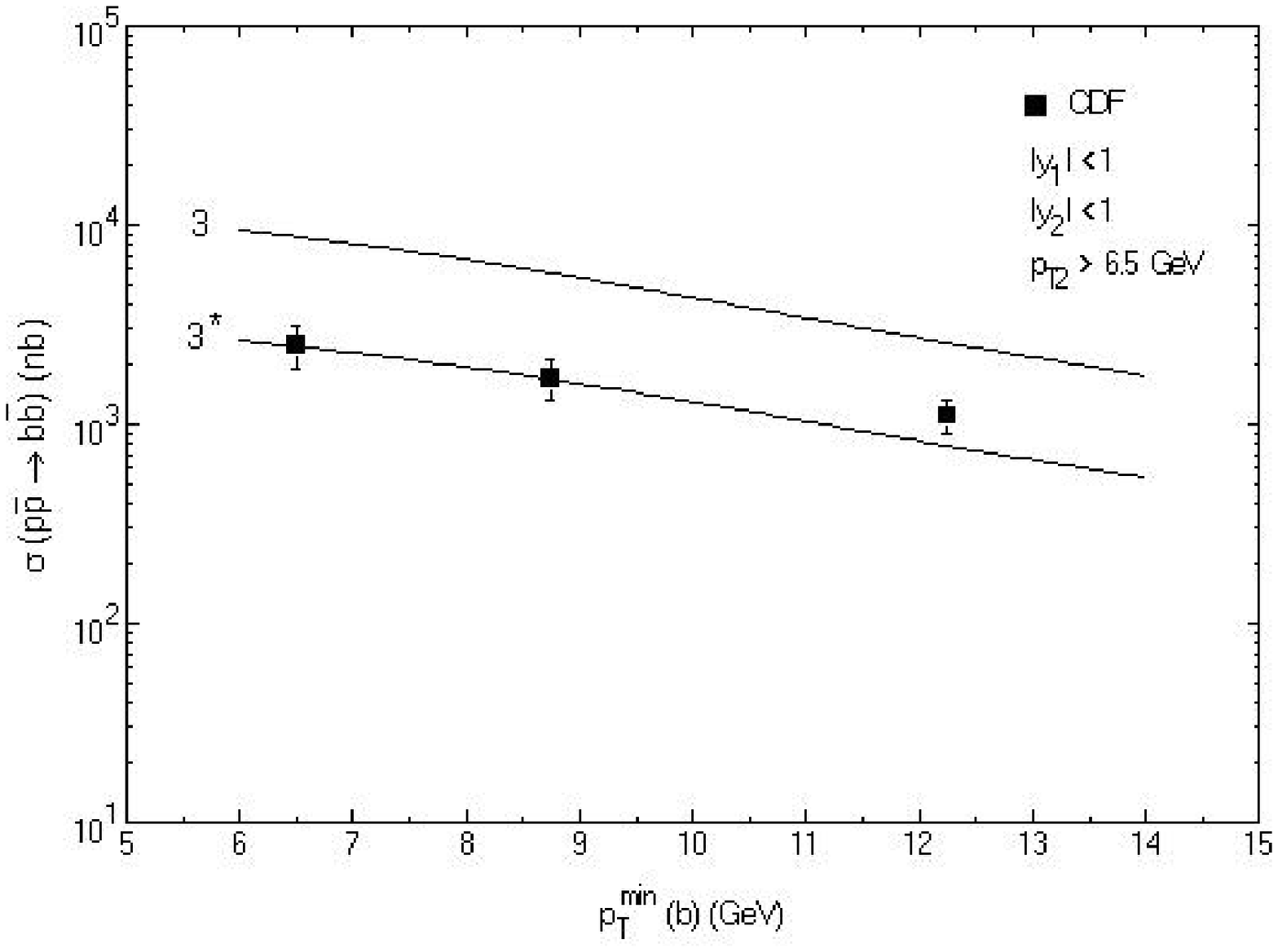,width=10cm,height=7cm}
\end{center}
\caption{   
 The $b\bar b$ production cross secton at $\sqrt s$ = 1.8 TeV.
Curves 3 and 3$^*$ are the same as in Fig. 6.}
\label{fig7}
\end{figure}
In Fig. 4, 5 the curves 1 correspond to the calculations in the framework
of the SPM  using the gluon density $xG(x,Q^2)$ 
from the standard GRV set~\cite{r27}. The curves 2, 3, 4, 5 are the
results of calculations in the framework of the SHA  using
the unintegrated gluon distribution in RS ($Q_0^2 = 4$ GeV$^2$),
LRSS ($Q_0^2 = 2$ GeV$^2$), BFKL ($Q_0^2 = 1$ GeV$^2$) parametrizations
 and the one (22) ($Q_0^2 = 1$ GeV$^2$) accordingly.
\begin{figure}
\begin{center}
\epsfig{figure= 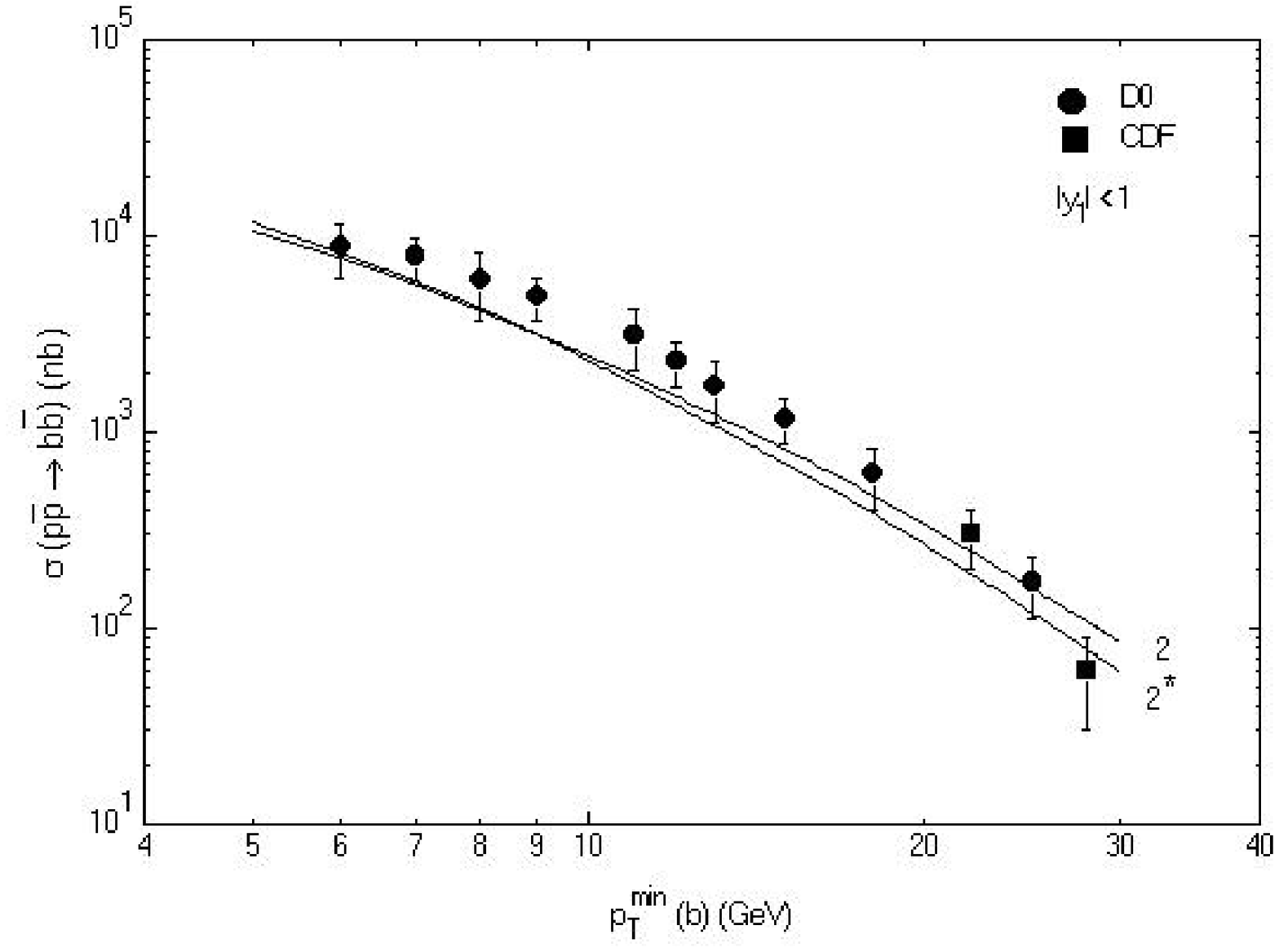,width=10cm,height=7cm}
\end{center}
\caption{   
 The $b\bar b$ production cross secton at $\sqrt s$ = 1.8 TeV.
Curves 2 and 2$^*$ correspond to the SHA calculations with
the RS parametrization of unintegrated gluon distribution
using the matrix elements (22) and the ones from ref.~\cite
{r6}.}
\label{fig8}
\end{figure}

 One can see that the SHA
curves 2, 4 and 5 describe the D0 and CDF experimental data very well
in comparison with the SPM results (curves 1).
 The LRRS parameterization of the unintegrated gluon distribution 
overestimates the  experimental $b\bar b$ production cross section
~\cite{r1,r2}.

We tried to describe these experimental data with help the LRSS 
paremetrization fitting the $Q_0^2$, $m_b$ and $\Lambda_{QCD}$
parameters in region: 1 GeV$^2 \le Q_0^2 \le 4$ GeV$^2$, 4.5 GeV $\le m_b 
\le 5.0$ GeV and 100 MeV $\le \Lambda_{QCD} \le 250$ MeV.
We obtained well agreement of the theoretical results (curves 3$^*$
in Figs. 6 and 7) 
with the TEVATRON experimantal data at the following
 values of the parameters: $Q^2_0 = 4$ GeV$^2$, $m_b = 5.0$ GeV
and $\Lambda_{QCD} = 100$ MeV~\footnote{The experimental data
for $b\bar b$ photoproduction obtained by the H1 and ZEUS Collaborations
~\cite{r34,r35} were described in the framework of the SHA with
LRSS parametrization of the unintegrated gluon distribution
at $Q^2_0 = 2$ GeV$^2$ only (at $m_b = 4.75$ GeV and $\Lambda_{QCD} = 200$
MeV)\cite{r9}.}.

In Fig. 8 we show the results obtained  using the off mass shell 
matrix elements of gluon-gluon fusion subprocesses in the form (22) and 
the ones from ref.~\cite{r6} with using  the RS parametrization 
of  unintegrated gluon distribution as example.
We see that these results coincide very well.

\section{Conclusions}

We considered the process of
inelastic heavy quark production at TEVATRON  in  the framework of the
semihard QCD approach with emphasis on the BFKL dynamics of gluon 
distributions. We investigated the cross section
$\sigma (p_T > p_T^{min})$ of inelastic
$b-$quark hadroproduction as a function of 
different unintegrated gluon distributions. It is shown that 
the description of the $b-$quark inelastic
cross section at TEVATRON energies is achieved\footnote
{When the present paper has been prepared to publication
the ref.~\cite{r36,r37} appeared, in which it was obtained
the similar results in the framework of the $k_T$
factorization approach.}  in the cases
 of the RS, BFKL parameterizations and also the
parametrization (22) at realistic values of QCD
parameters ($m_b = 4.75$ GeV and $\Lambda_{QCD} = 150$ MeV).
The LRSS parametrization describes the D0 and CDF experimental
data with another values of parameters 
( $m_b = 5.0$ GeV and $\Lambda_{QCD} = 100$ MeV) in
comparison with the description of the $b\bar b$ quark photoproduction 
data at HERA in the framework of the SHA~\cite{r9}.

 One of the main goals of our investigations consist of  a search
in the framework of the SHA
of the "universal" unintegrated gluon distribution.
In our opinion the studies of many high energy heavy
quark production processes in the framework of the SHA
~\cite{r9,r12,r13,r14,r17}
have showed that the so called BFKL unintegrated gluon distribution
function~\cite{r26} is one of the candidates for a role of 
universal gluon distribution.

\section{Acknowledgments}

 One of us (N.Z.) would like to thank S. Baranov, H. Jung and 
 L. J\"onsson
 for numerous discussions of different aspects of the QCD SHA,
  the Elementary Particle Department of Institute of Physics
of Lund University for hospitality.
This work has been supported in part by the Royal Swedish
Academy of Sciences.
 N.Z. thanks also I. Korzhavina for careful reading
the manuscript and useful remarks.

\end{document}